\documentclass[runningheads,a4paper]{llncs}

\usepackage{amssymb}
\usepackage{amsmath}
\usepackage{graphicx}
\usepackage{enumitem}
\usepackage{url}
\urldef{\mailsa}\path|petra.surynkova@jku.at|
\newcommand{\keywords}[1]{\par\addvspace\baselineskip
\noindent\keywordname\enspace\ignorespaces#1}

\makeindex
\begin{document}

\mainmatter  % start of an individual contribution

\title{Selecting the Best Quadrilateral Mesh for Given Planar Shape}

\titlerunning{Selecting the Best Quadrilateral Mesh}

\author{Petra Surynkov\'{a}}
\authorrunning{Selecting the Best Quadrilateral Mesh}
% (feature abused for this document to repeat the title also on left hand pages)

% the affiliations are given next; don't give your e-mail address
% unless you accept that it will be published
\institute{Johannes Kepler University Linz,\\
Altenberger Str. 69, 4040 Linz, Austria\\
\mailsa\\}

%
% NB: a more complex sample for affiliations and the mapping to the
% corresponding authors can be found in the file "llncs.dem"
% (search for the string "\mainmatter" where a contribution starts).
% "llncs.dem" accompanies the document class "llncs.cls".
%

\maketitle

\begin{abstract}
The problem of mesh matching is addressed in this work. For a given $n$-sided planar region bounded by one loop of $n$ polylines we are selecting optimal quadrilateral mesh from existing catalogue of meshes. The formulation of matching between planar shape and quadrilateral mesh from the catalogue is based on the problem of finding longest common subsequence (LCS). Theoretical foundation of mesh matching method is provided. Suggested method represents a viable technique for selecting best mesh for planar region and stepping stone for further parametrization of the region.
\end{abstract}

\section{Introduction and Motivation}

Finding a quadrilateral mesh that matches a given shape - a so called mesh matching problem - represents an important problem in computer aided design, \cite{BoLe13}, \cite{Ed01}, \cite{KaNi07}, and \cite{NaSa09}. The task is specified by user input which describes planar shape to be covered with quadrilateral mesh. In this work we consider only simply connected planar domains which need to be filled by quadrilateral mesh. The planar shape is represented by its boundary drawn in plane typically. In addition to this, the size of mesh, that is, the number of its vertices, is specified by the user as well.

\begin{figure*}[htb]\centering
	\includegraphics{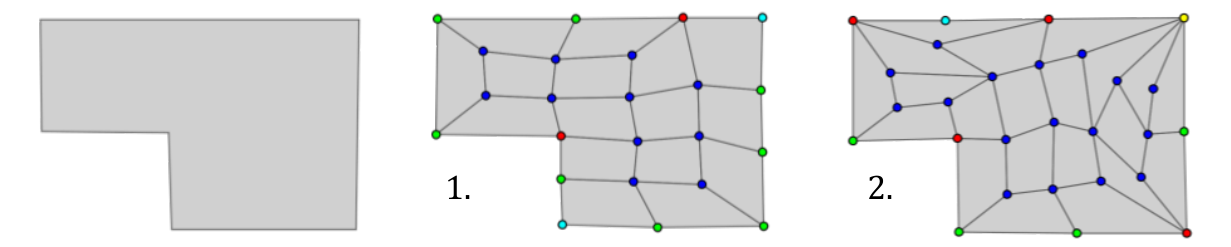}
	\caption{Two examples of quadrilateral meshes for user input on the left. The first mesh expresses the shape of user input better than the second mesh. First mesh has lower valences of boundary vertices in {\em convex} vertices of user input and higher valences in {\em concave} vertices. The various valences of boundary vertices have different color.}
	\label{fig:motivation01}
\end{figure*}

There are many ways how to cover the given shape with a mesh as even single mesh can be matched to the shape in many ways. Moreover there are many non-isomorphic meshes of given number of vertices which makes the problem of finding appropriate mesh for given shape hard. An example of two matchings  of different quality between user input and quadrilateral meshes is provided in Figure \ref{fig:motivation01}.

Trying to find a suitable matching between the shape and a mesh in an intuitive way by evaluating all the possible coverings would quickly lead to combinatorial explosion as there is exponential number of such covering. We were trying to mitigate this difficulty by finding an appropriate concept from computer science that can be used for reasoning about the mesh matching problem in a more efficient way. Our finding is that the concept of \textit{longest common subsequence} (LCS) \cite{BeHa00}, \cite{Hi77} can be employed in mesh matching problem to avoid the combinatorial explosion. Given an objective function for measuring quality of matching, the suitable application of the LCS algorithm would automatically prune out coverings of the input shape by mesh that have no chance to maximize the objective function. Hence instead of dealing with the exponential number of coverings the algorithm goes directly to the optimal one in polynomial number of steps.

In this article we for the first time describe concept of mesh matching to a given planar shape in the formal mathematical way. Having the precise mathematical formulation of the problem we could describe objective function that corresponds to the quality of matching. We also designed a computationally efficient method for mesh selection which is optimal with respect to the matching quality objective. Our method generalizes the common problem of LCS. In contrast to the original LCS, where only the relation of equality between symbols is considered, we allow more general relations between symbols that reflects various cases that arise in mesh matching problem.

\section{Background}
A quadrilateral mesh, \cite{BoKo10}, \cite{Mit96}, \cite{SuJu16}  is a triple $(V, E, Q)$ where $V$ is a set of vertices, $E$ is a set of edges, and $Q$ is a set of quadrilaterals. There exists an embedding of $(V, E, Q)$ into 2D plane such that each vertex is assigned a point in the plane and each edge is assigned a curve in the plane, so that curves connect vertices and each quad is represented in the plane as a quadrilateral. In our study we furthermore assume only quadrilateral meshes that form a connected, conforming (i.e. free from T-junctions), orientable \textit{2D manifold with boundary}, \cite{BoKo10}, \cite{LuRe03}, i.e. our quadrilateral meshes are defined for segmentation of simply connected planar domains. 
	
Regarding the terminology, an edge of the mesh with two incident quads is said to be \textit{internal}, while an edge with just one incident quad is said to be \textit{boundary}. A vertex of an internal edge is also said to be \textit{internal}, otherwise it is said to be \textit{boundary}. The valence of a vertex is the number of edges incident to that vertex.

%which we denote by $\ell(v)$,

\subsection{Existing Catalogues of Meshes}
Our work considers the existing context of literature and software in geometry and computer aided design. Extensive works on providing catalogue of meshes of various sizes had been done previously, \cite{BuJu16},  \cite{SuJu16}. Our approach in this work is to integrate our mesh selection method with existing mesh catalogue that is kept as a database generated by procedural algorithm - that is, our method will select the optimal mesh out of the catalogue of meshes.

The catalogue of meshes which we use as the input for our selection method consists of quadrilateral meshes of certain class and was generated in the previous work, \cite{SuJu16}. Without constraints, the number of possible meshes is too high. Thus, we consider only valences $\leq 5$ for the both boundary and internal vertices which represents standard restriction for numerical applications \cite{LiXi11} and what is used in related literature \cite{PeBa14}, \cite{TaPa14}. 

The quadrilateral meshes in the catalogue furthermore satisfy the following invariant:

\begin{enumerate}[label=(\Roman*)]
	\item \label{itm:invariant} At least one vertex of each internal edge is internal.
\end{enumerate}

Procedural mesh catalogue generates exponential number of meshes with respect 
to the number of internal vertices of mesh, i.e. the number of internal vertices is specified as the input. Hence, our mesh selection method should make a consideration about matching of a single mesh very quickly to be able to find suitable mesh for given shape in reasonable time. More precisely, we cannot afford to perform any kind of exponential time search or other time consuming operation.

\section{Longest Common Subsequence Problem}
The most simplified version of the longest common subsequence problem consists in finding common subsequence within given two sequences of symbols. Informally said, the task is to delete some symbols in given two sequences so that the resulting sequences - called common subsequences - will be the same. The objective is to obtain as long as possible sequences at the end (in other words we want to make as few as possible symbol deletions) - that is, longest common subsequences.

Consider a simple examle of two strings (set of symbols correspond to latin alphabet) {\tt"alpha"} and {\tt"aleph"}. These two sequences are obviously different but after deleting last {\tt'a'} in the first sequence and middle {\tt'e'} in the second string we obtain {\tt"alph"} in both cases which is the longest common subsequence for this example.

The very positive aspect about the LCS problem is that a variety of efficient (polynomial time) algorithms exist that solve this optimization problem, \cite{Uk85}, \cite{UlAh76}.

\subsection{Application in Shape Matching}

Low time complexity makes LCS algorithms good candidates for using them as a basis for consideration about mesh matching. However, mesh matching problem and LCS problem are completely different concepts hence we need to show first what are the similarities between these problems.

When a user specifies his planar shape we can regard his input as an abstract information that can be annotated by a sequence of symbols. Information about the {\em length} of edges of the boundary, which vertices on the boundary are {\em convex}, which are {\em concave} can be read from the input. Such information can be encoded into a sequence of symbols.

At the same time, we need to annotate boundaries of meshes stored in the catalogue using a sequence of symbols in correspondence with annotation of inputs. There is considerable discrepancy between user inputs in the form of planar shapes and representation of meshes in existing catalogues - most catalogues represent meshes in the abstract form as list of quadrilaterals and their interconnections. Moreover we need to reflect certain level of flexibility of mesh matching with respect to given shape - a mesh may be matched to the shape in a not ideal way while no better matching is possible.

We therefore introduce {\em lattice} of classification of mesh boundary vertices with respect to properties of {\em convexity}, {\em concavity}, and {\em others}. Unlike in the case of standard LCS where we compare a pair of symbols whether they are same or not, here we are more flexible. When we compare a vertex from user input shape with a boundary vertex of mesh from the catalogue, the quality of matching between these two vertices is determined by the {\em lattice}. For example if a {\em convex} point from user input is being compared with a vertex from mesh, then {\em lattice} for classification of {\em convex} vertices is used to determine the quality of correspondence between the two (the {\em lattice} thus provide classification from complete match between {\em convex} and {\em convex} point to complete mismatch between {\em convex} and {\em concave} point).

This operation of comparison between a point from user input and a boundary vertex of a mesh comes as a parameter to an LCS algorithm instead of standard equality between symbols. The output of the algorithm hence will be pair of sequences whose symbols at corresponding positions represent best possible match according to the {\em lattice}. When this output is interpreted back to the world of geometry we have an optimal matching of a given mesh from catalogue to the given user input. Hence we are able to select optimal mesh from the catalogue provided that consideration about the single mesh is fast enough. 

\begin{figure*}[htb]\centering
	\includegraphics{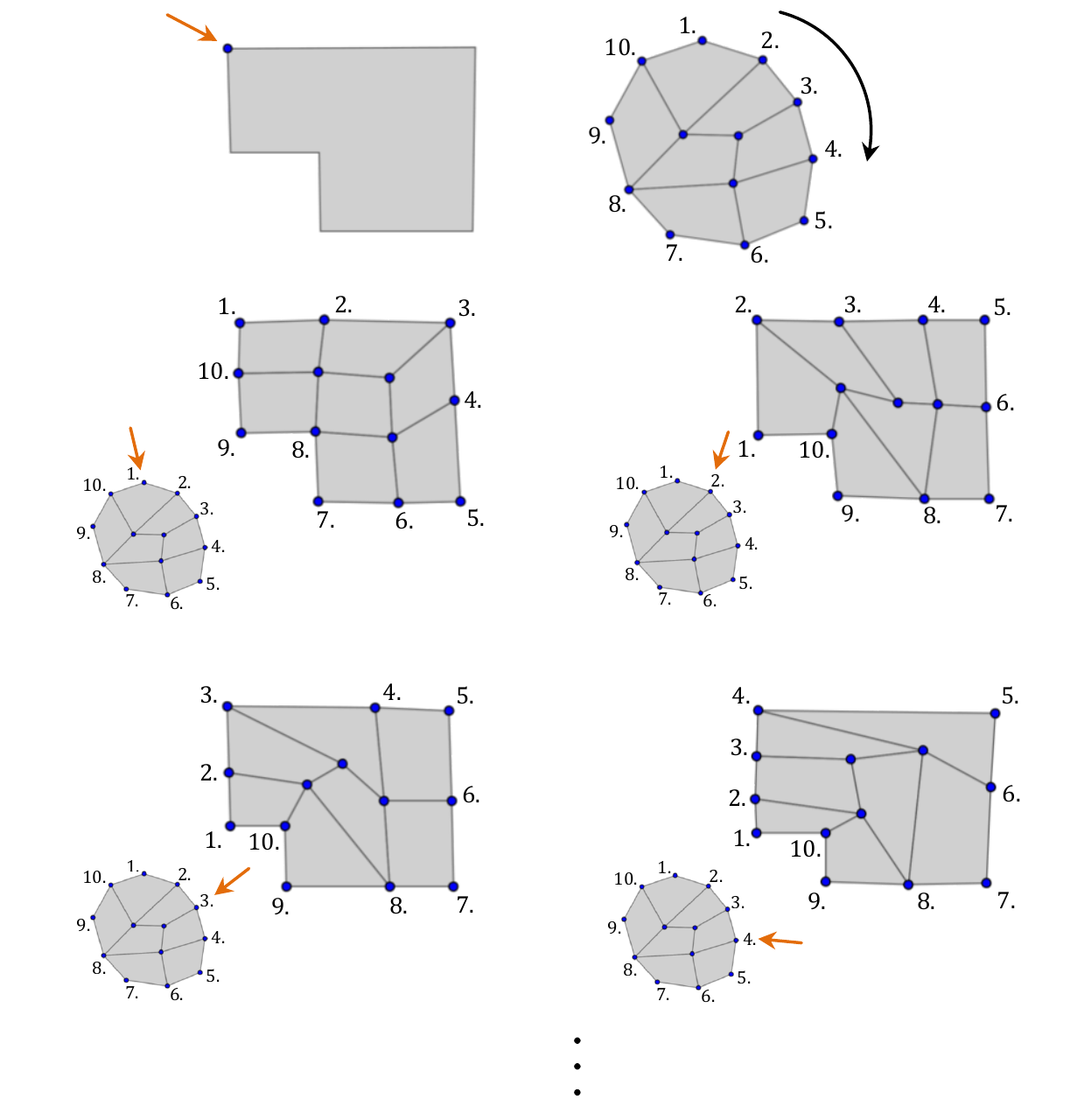}
	\caption{An example of mesh matching between one quadrilateral mesh from the catalogue and user input. Point in the user input is fixed and the rotations of mesh are considered.}
	\label{fig:matching}
\end{figure*}

The method works for fixed starting point in a testing quadrilateral mesh which is compared to fixed point in the given user input. For finding an optimal matching between the user input and a mesh from the catalogue we consider all rotation of a testing mesh, i.e. for fixed point in the user input we test all possible starting points in a mesh. This approach is illustrated in Figure \ref{fig:matching}.

\section{Optimal Mesh Selection Method}
\begin{figure*}[htb]\centering
	\includegraphics{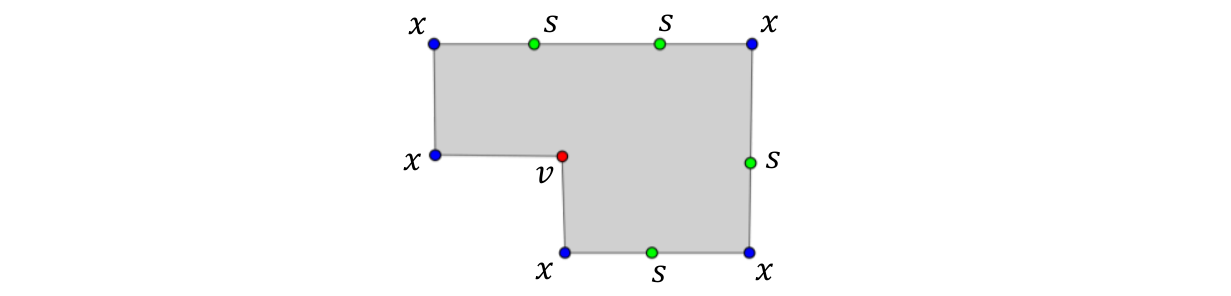}
	\caption{Three types of boundary vertices in the given user input.}
	\label{fig:categorization}
\end{figure*}

Our basic mesh matching method assumes a catalogue of meshes represented as a list of interconnected quadrilaterals. The user input given as a planar shape is further processed and is assigned a sequence of boundary points. We distinguish three types of boundary vertices, see Figure \ref{fig:categorization}:

\begin{itemize}
	\item{{\bf straight} point} - denoted by symbol $s$
	\item{{\bf convex} point} - denoted by symbol $x$
	\item{{\bf concave} point} - denoted by symbol $v$
\end{itemize}

{\em Convex} and {\em concave} points are extracted naturally from the user defined input planar shape. {\em Straight} points are assigned to lines of the boundary of shape. That is, each line is assigned certain number of internal {\em straight} points according to its length. Longer boundary line have more internal {\em straight} vertices.

As meshes in the catalogue are represented in the abstract way there is not much information available from which a corresponding annotation can be constructed. The only usable information about meshes are valences of their boundary vertices which in case of quadrilateral meshes are from the range ${2,3,4,5}$. To further increase amount of information about a given boundary vertex we also consider valences of its neighbors which allows us to determine which vertex is more likely {\em convex} or which vertex is more likely {\em concave}.

\begin{figure*}[htb]\centering
	\includegraphics{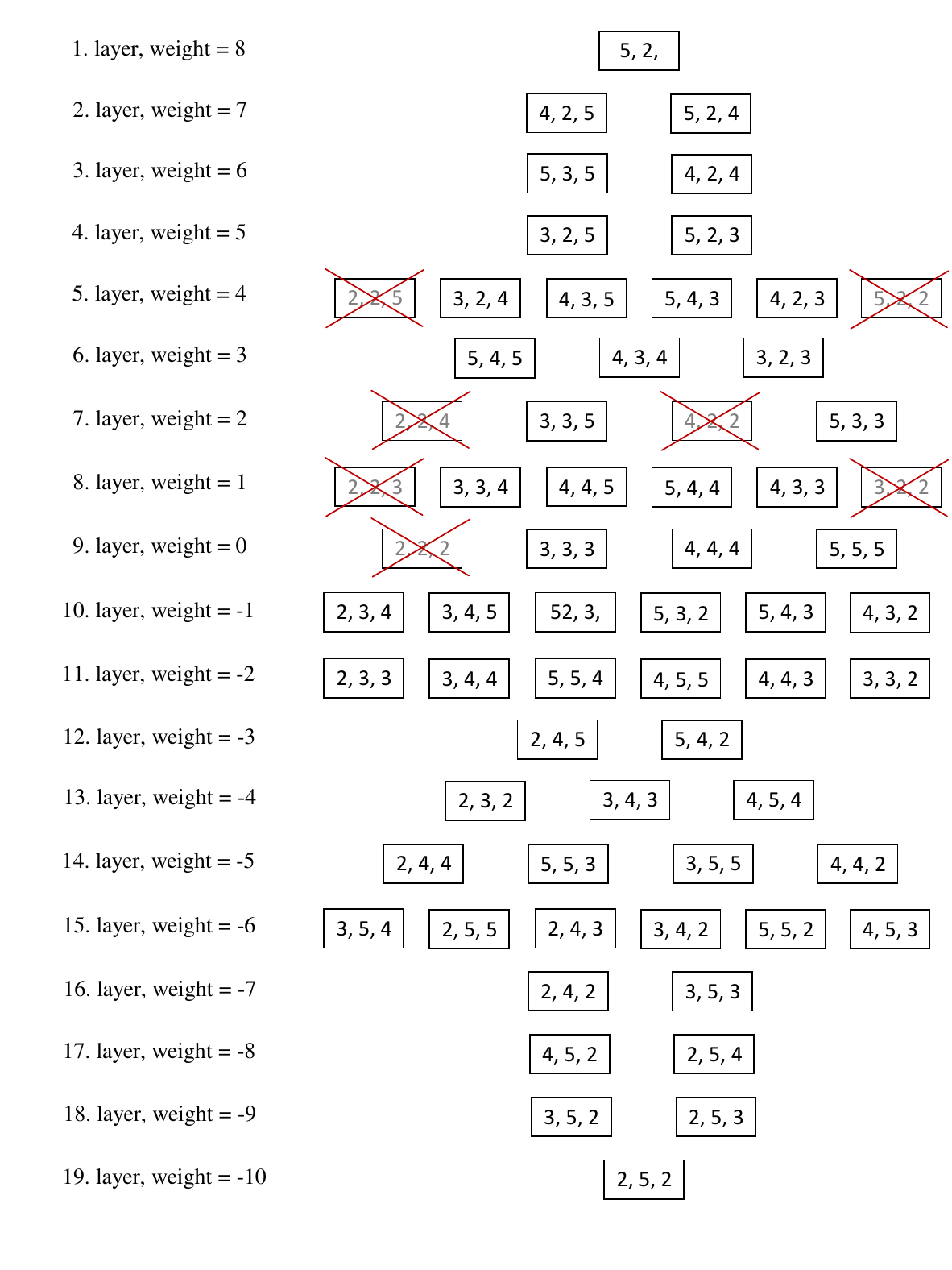}
	\caption{Suggested lattice for {\em convex} point in the given user input.}
	\label{fig:lattice_convex}
\end{figure*}

Intuitively {\em convex} boundary vertices have low valence while their neighbors have high valence. To be able to formalize this likeliness to be {\em straight}, {\em convex}, or {\em concave} point we introduce a lattice for each type of point. Figure \ref{fig:lattice_convex} shows a suggested lattice for {\em convex} point in the given user input. Some triples of valences of boundary points in the lattice are weeded out because they cannot occur in our certain class of meshes. The lattices for {\em concave} and {\em straight} points are designed analogically.

The higher the vertex is classified within the lattice the more likely it can be matched to a {\em convex} point on the user input. The design of lattice is a matter of careful consideration of visual appearance of matching and experience of the expert designer. Furthermore, each level of lattice is assigned an integer weight that will be reflected by the modified LCS algorithm when comparing symbol from the input sequence with mesh boundary vertex. Positive weights represent likeliness of match between points while negative values stand for likeliness of mismatch. Absolute value of the weight represent measure of match or mismatch. A special weight $-\infty$ is reserved to denote complete mismatch between the pair of vertices (this corresponds to disequality between symbols in the standard LCS algorithm).

Our modified LCS algorithm assumes two input sequences of symbols - the first from user input; the second obtained by annotating mesh from the catalogue. The symbols of the second sequence have the following format of triples: $[u,v,w]$ where $v \in {2,3,4,5}$ is a valence of mesh boundary vertex, and $u,w \in {2,3,4,5}$ are valences of counter-clock-wise and clock-wise neighbors of $v$ respectively.

\begin{figure*}[htb]\centering
	\includegraphics{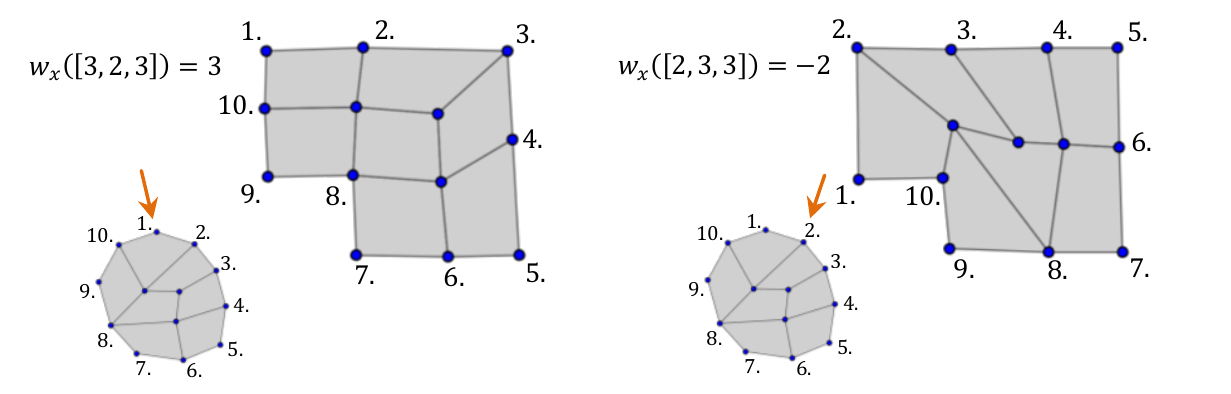}
	\caption{An example of weights of selected mesh boundary vertex with respect to the fine-grained lattice for {\em convex} point.}
	\label{fig:example}
\end{figure*}

Weights are assigned to triples$[u,v,w]$ by weight functions. Formally we introduce a weight function for each type of point. That is, we have functions:

\[ w_s, w_x, w_v:\{2,3,4,5\}^3  \rightarrow \mathbb{Z} \cup \{-\infty\} \]

which assigns triples of valences their weights with respect to the lattice for {\em straight}, {\em convex}, and {\em concave} point, respectively, see an example in Figure \ref{fig:example}. These weight fuctions are used by the modified LCS algoritm when it is making comparing a pair of symbols (first one from the first sequnce second one from the second sequence).

We omit implementation details of the modified LCS algorithm for the sake of brevity. However, important high level specialty of our version of LCS is that it assumes the length of the first sequence to be at least the length of the second sequence. Moreover, deletions of symbols can be made from the first sequence only (from the user input). It is natural assumption as we want to match all the vertices of the catalogue mesh to some point in the user defined shape.

The two sequences of the same length are valuated by a utility which is calculated as the sum of weights of symbols from the second sequence with respect to lattices corresponding to respective symbols from the first sequence. The task of our modified LCS algorithm is to compute longest common subsequence out of given user defined input sequence (the first sequence) with the highest possible utility.

Formally we define overall utility function of a pair of sequences of symbols discussed above as follows 

\[ f([s_1,...s_n],[[u_1,v_1,w_1],...[u_n,v_n,w_n]])  = \sum_{i=1}^{n} w(s_i,[u_i,v_i,w_i]) \]

where

$$
w(s_i,[u_i,v_i,w_i]) =
  \begin{cases}
    w_s([u_i,v_i,w_i]),  & \quad \mbox{if }  s_i=\mathtt{s} \\ 
    w_x([u_i,v_i,w_i),   & \quad \mbox{if } s_i=\mathtt{x} \\
    w_v([u_i,v_i,w_i)    & \quad \mbox{if } s_i=\mathtt{v} \\
  \end{cases}
$$

The algorithm finds a sequence $[s_1,...s_n]$ for the input pair of sequences $[s'_1,...s'_m]$ and $[[u_1,v_1,w_1],...[u_n,v_n,w_n]$ where $n \leq m$ so that $[s_1,...s_n]$ is a subsequence of $[s'_1,...s'_m]$ and its  $f$ value is maximum.

\subsection{Design of Lattices}
\begin{figure*}[htb]\centering
	\includegraphics{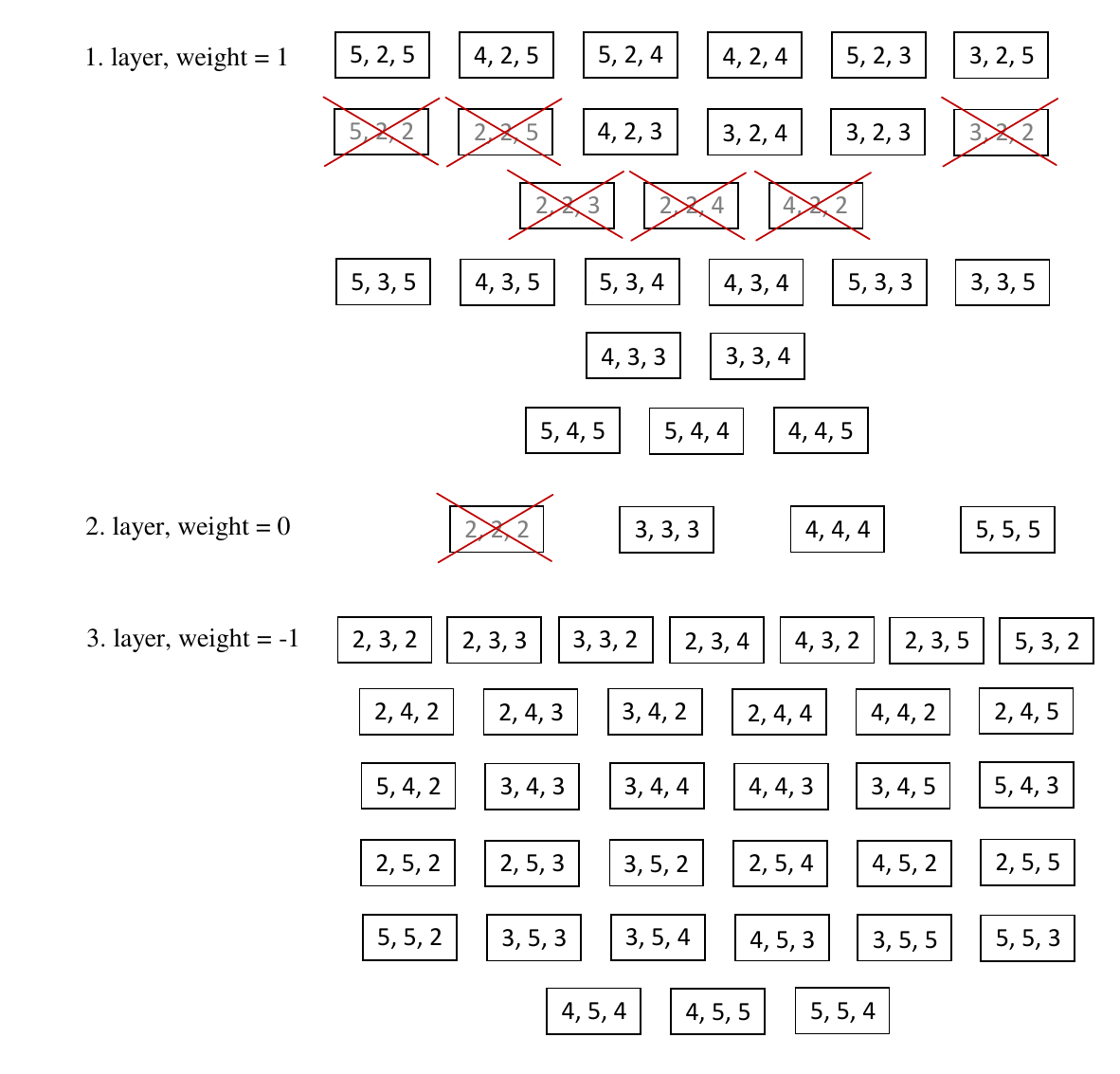}
	\caption{An example of coarse lattice for {\em convex} point in the given user input.}
	\label{fig:coarse_convex}
\end{figure*}

\begin{figure*}[htb]\centering
	\includegraphics{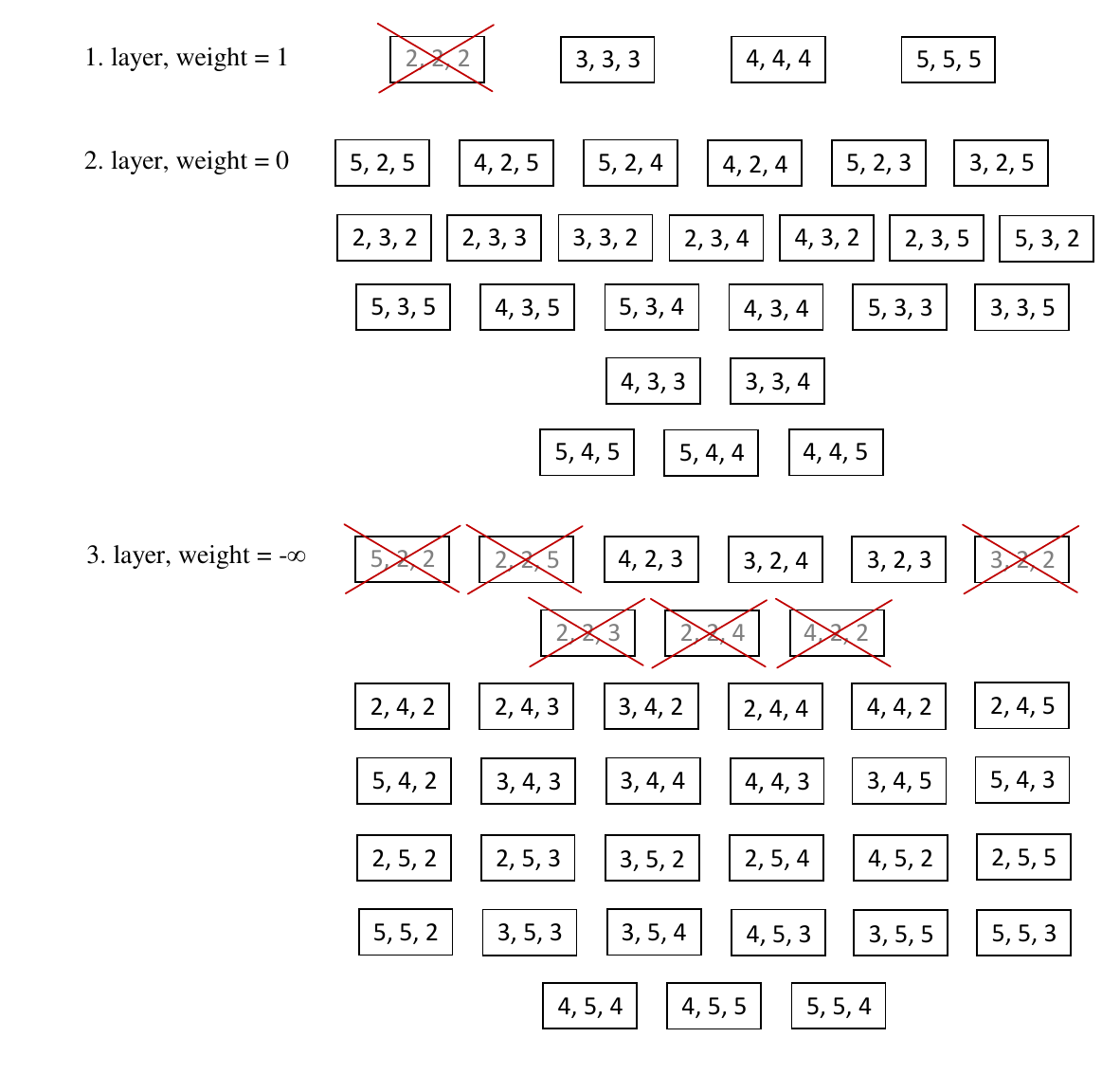}
	\caption{An example of coarse lattice for {\em straight} point in the given user input.}
	\label{fig:coarse_straight}
\end{figure*}

The design of {\em lattice} is a matter of experience of the expert designer. We suggested a lattice for each type of point in the user input. Our lattices are fine-grained so they enable to evaluate mesh matching more precisely than coarse lattices. An example of coarse lattice for {\em convex} point is shown in Figure \ref{fig:coarse_convex} and an example of coarse lattice for {\em straight} point is shown in Figure \ref{fig:coarse_straight}.

Note that occurrence of $-\infty$ represents strict impossibility to match a point of given type from user input (corresponds to the type of lattice) to a vertex from a mesh. If the LCS algorithm cannot find a correspondence between point from the user input and some mesh vertex so that utility other than $-\infty$ is assigned to that correspondence the point must be treated as deleted by the LCS algorithm. This is the only case when LCS performs deletion from some of its input sequences.  

\subsection{Properties of the Method}

Clearly as LCS algorithms are optimal in their nature our modified algorithm enables finding optimal match between used defined shape and mesh from the catalogue with respect to given objective function optimally. Moreover the algorithm requires polynomial time and space which makes it an excellent candidate for selecting a mesh with the best match for the given user input.

\section{Conclusions and Future Work}
A new approach for finding a quadrilateral mesh from the catalogue of quadrilateral meshes of certain class that matches a given user input was described in this work. The method conceptually builds on the known problem of finding longest common subsequence (LCS). As LCS and mesh matching are fundamentally different problems we proposed a series of techniques that allow us to transform mesh matching problem to LCS. These techniques include adaptation of the LCS algorithm and introduction of symbol comparison based on lattices that model likelihood of correspondence between user input points and mesh vertices.

The theoretical foundation of a method is provided. The major contribution is that viewing mesh matching problem through the concept of LCS allows mitigating the combinatorial complexity. Instead of evaluating all possible matchings between the user input and a mesh from the catalogue in the exponential time, the adapted LCS algorithm rules out partial matchings as early as possible if they turn out not to be optimal which leads eventually to the polynomial time.

In the future work we will focus on experimental evaluation which will be targeted on visual comparison of match matching obtained using fine-grained and coarse lattices. The interconnection of the mesh matching algorithm and the procedural catalogue is planned too.

Another interesting topic for the future work is to develop new techniques for evaluation best mesh according to the distribution of internal vertices.

\bibliography{reference}

\end{document}